\documentclass[a4paper]{jpconf}
\usepackage{graphicx}
\usepackage{amsmath}

\begin{document}
\title{Efficient charged particle propagation methods}

\author{P. Reichherzer$^{*1,2,3}$ and J. Becker Tjus$^{1,2}$}
\address{$^{1}$\quad Ruhr-Universit\"at Bochum, TP IV, Universit\"atsstrasse 150, 44801 Bochum, Germany}
\address{$^{2}$ \quad Ruhr Astroparticle and Plasma Physics Center (RAPP Center), 44780 Bochum, Germany}
\address{$^{3}$ \quad IRFU, CEA, Université Paris-Saclay, F-91191 Gif-sur-Yvette, France}

\ead{* patrick.reichherzer@ruhr-uni-bochum.de}

\begin{abstract}
In astrophysics, the search for sources of the highest-energy cosmic rays continues. For further progress, not only ever better observatories but also ever more realistic numerical simulations are needed. 
We compare different approaches for numerical test simulations of UHECRs in the IGMF and show that all methods provide correct statistical propagation characteristics of the particles in means of their diffusive behaviour. Through convergence tests, we show that the necessary requirements for the methods differ and ultimately reveal significant differences in the required simulation time.
\end{abstract}

\section{Introduction}\label{sec:1}
A key task in astrophysics is the search for sources of high-energy cosmic rays (CRs) \cite{Tjus2020}. Unlike uncharged messengers, such as photons and neutrinos, CRs are deflected on their journey to Earth due to their electrical charge in the intergalactic magnetic field (IGMF) and the Galactic magnetic field (GMF). These deflections across astronomical distances lead to an isotropization of arrival detections of CRs a loss of knowledge about their origin. 

In the search for the sources of CRs, not only the multimessenger signatures of known source types must be properly modeled, but also the transport properties of the charged particles in the sources \cite{BeckerTjus:2022} and on the way from the sources to Earth must be understood.

In this study, we investigate different propagation methods to study the transport of charged particles in magnetic fields. Section~\ref{sec:2} provides an overview of the proven methods that either solve the equation of motion (EOM) or are based on solving the diffusion equation. Furthermore, we present an approach based on the correlated random walk. Comparisons of these different propagation methods in an astrophysical context are presented in Section~\ref{sec:3}.

\section{Different Propagation Methods for Charged Particles in Magnetic Fields}\label{sec:2}

\subsection{Solving the Equation of Motion}
Particle trajectories in magnetic fields can be computed by solving the Lorentz equation
\begin{equation}\label{lorentz}
    \frac{\mathrm d \textbf{{v}}}{\mathrm{d} t} = \frac{q}{mc}(\textbf{v} \times \textbf{B}),
\end{equation}
with the mass $m$, speed $c$, particle velocity $\textbf{v}$, magnetic field $\textbf{B}$, and charge $q$ of the particle. This first order differential euqation can be numerically solved using many different algorithms. The Boris-Push (BP) method \cite{boris1972proceedings} and the Runge-Kutta type Cash-Karp (CK) method \cite{CashKarp1990} are two well-tested ones \cite{reichherzer2021:b}. 
Both algorithms support adaptive step sizes. Note that with the EOM approach, the local magnetic field vector must be known for each integration step.

\subsubsection{Synthetic Turbulence Generation}
Synthetic turbulence can be generated via the summation of planar waves with different wave numbers, amplitudes, and directions, by either
\begin{enumerate}
    \item storing discrete magnetic field vectors of the turbulence on a large grid with $N_\mathrm{grid}^3$ grid points by using an inverse discrete Fourier transform. During run-time, the local magnetic field is computed via interpolation of the surrounding grid points (see e.g. \cite{reichherzer2020, Reichherzer:2021a}),
    \item or by the summation of different amplitudes, wavenumbers, and directions during run-time at the exact position where it is needed (\textit{plane-wave} (PW) \textit{turbulence}) \cite{Giacalone1999, TD13}.
\end{enumerate}
See \cite{Schlegel2020} for a detailed comparison.

\subsubsection{Particle Trajectory Integration}
\paragraph{Runge-Kutta Method}
A first order differential euqation $\mathrm{d} y / \mathrm{d} t = f(t,y)$ can be numerically solved with an embedded fifth order Runge-Kutta method.
Comparison of this fifth-order formula with different order Runga-Kutta expressions allows for an error estimation and a step size adaption for numerical error limitation. The CK method employs this error-limiting technique. This method is known for its good properties for charged particles in magnetic fields.
Note that the CK method is not energy-conserving. This issue can be fixed by
\begin{enumerate}
    \item manually enforcing conservation of energy, as it is done in CRPropa for example. Fixing the energy to a constant value requires adjusting the momentum components. 
    \item choosing a good resolution of the particle trajectory by using small step sizes. In this way, the local error can be kept small and an almost energy- and momentum-preserving behavior for the propagation of charged particles in the magnetic field can be ensured. This fix is accompanied by increased simulation times.
\end{enumerate}

\paragraph{Boris-Push Method}
In contrast to the CK algorithm, the BP method is intrinsically energy-conserving. The BP method applies a leapfrog integration scheme. In contrast to implicit methods, this explicit BP method, as well as the explicit CK method, only evaluates the Lorentz from the previous step, which requires smaller step sizes but improves the simulation speed. \\

\begin{figure}[htbp]
    \centering
    \includegraphics[width=0.75\textwidth]{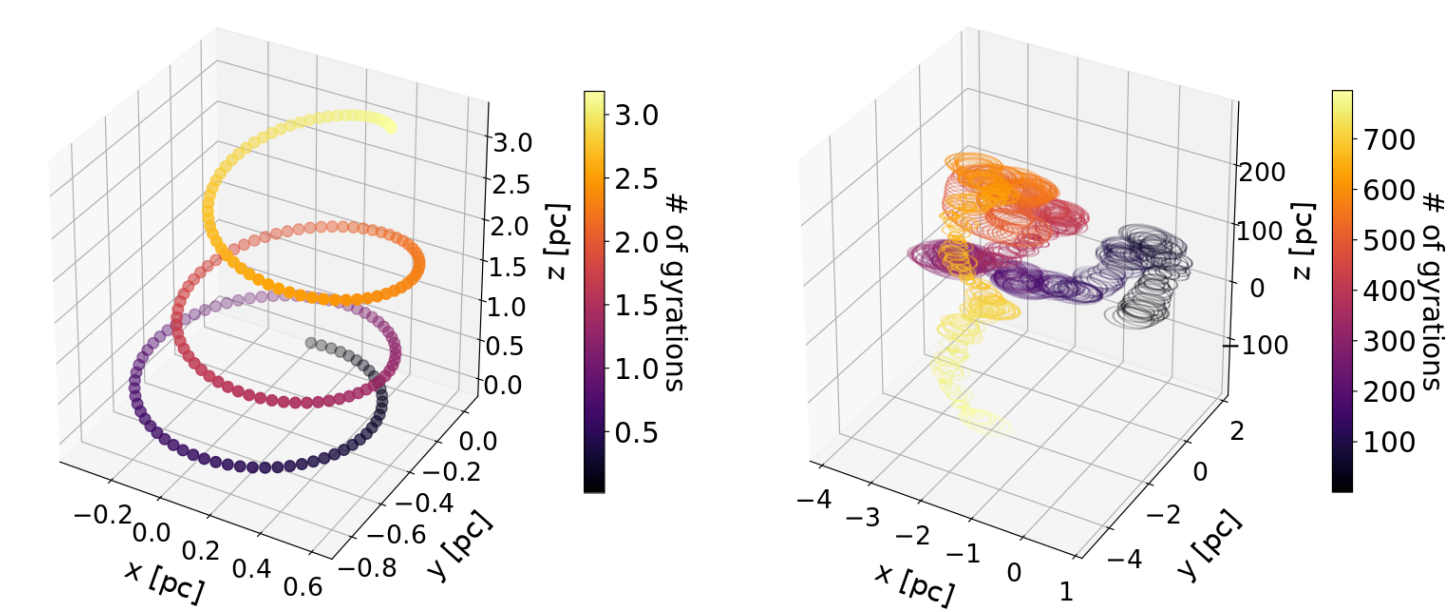}
    \caption{Charged particle trajectory in turbulence plus an ordered magnetic field along the $z$-axis simulated with an EOM approach. The first gyrations (left) show the initial ballistic propagation character, while the diffusive behavior dominates in the limit of large times (right). Taken from \cite{masterthesis}. 
    \label{fig:trajectory} 
    }
\end{figure}

Figure~\ref{fig:trajectory} shows the trajectory of a charged particle in a directed background magnetic field along the $z$-axis plus a turbulent component. In the left panel, the local gyration movement caused by the background field becomes visible. In the right panel, on the other hand, the chaotic character due to the turbulent component of the magnetic field, which causes a random walk trajectory of the charged particle, becomes clear at large times.

\subsection{Solving the Transport Equation}
The dynamics of particle transport can be described statistically in the limit of large times via a Fokker-Planck transport equation
\begin{equation}
    \frac{\partial n}{\partial t} + \textbf{u} \cdot \nabla n = \nabla \cdot (\hat{\kappa} \nabla n) + \frac{1}{p^2}\frac{\partial}{\partial p} \left(p^2 \kappa_p \frac{\partial n}{\partial p} \right) + \frac{p}{3} (\nabla \cdot \textbf{u}) \frac{\partial n}{\partial p} + S,
\end{equation}
with the isotropic particle density $n$, the advection speed $\textbf{u}$, the spatial diffusion tensor $\hat{\kappa}$, the momentum diffusion coefficient $\kappa_p$, and $S$ the source term.
The diagonal elements $\kappa_{ii} = \lim\limits_{t \rightarrow \infty}{\langle (\Delta x_i)^2 \rangle}/2t$ of the spatial diffusion tensor are identical for turbulence without a guide field. Here, $i$ indicates the direction $x, y,$ or $z$ and the notation $\langle...\rangle$ refers to averaging over all particles. The transport equation can be solved numerically either
\begin{enumerate}
    \item discretized in real and momentum space (e.g. \cite{strong_propagation_1998, Dragon}),
    \item by transformation into stochastic differential equations \cite{CRPropa3_1}.
\end{enumerate}

\subsubsection{Diffusion Equation}

The most simple form of the transport equation is the diffusion equation that only assumes spatial diffusive transport in a steady-state:
\begin{equation}
    \frac{\partial n}{\partial t} = \sum_i \kappa_i \frac{\partial^2 n}{\partial x_i^2}.
\end{equation}

\subsubsection{Telegraph Equation}
In contrast to the diffusion equation, the telegraph equation takes into account the initial ballistic propagation phase by means of an additional term \cite{Tautz2016},
\begin{equation}\label{eq:telegraph}
\frac{\partial n}{\partial t} + \sum_i \tau_i \frac{\partial^2 n}{\partial t^2}= \sum_i \kappa_i \frac{\partial^2 n}{\partial x_i^2},
\end{equation}
where $\tau_i$ denotes the time scale for particles to become diffusive, and $\kappa_i$ is the diffusion coefficient, from which the relevant parameters of the correlated random walk (CRW) can be determined. This approach allows the correct description of the initial particle transport (left panel of Fig.~\ref{fig:trajectory}) and the steady-state diffusive regime (right panel of Fig.~\ref{fig:trajectory}).

\subsection{Correlated Random Walk}
The concept of CRW has been discussed in various contexts in the literature, such as when describing animal trails (see e.g., \cite{Codling2008}), but can also be applied for cosmic-ray propagation \cite{Seta2019}.

During the CRW, only two substeps are performed in each propagation step. The first substep determines if the particle changes its direction. The probability $\epsilon$ the change in direction can be derived from the diffusion coefficient. The second substep moves the particle along the direction established in the first substep for a time interval $\tau_\mathrm{s}$.

First, when considering only one dimension, a set of differential equations can be formulated, which describes the dynamics of the particle distribution pointing in the positive direction $\alpha$, and in the negative direction $\beta$, respectively,
\begin{equation}\label{eq:alpha}
\frac{\partial \alpha}{\partial t} = - v \frac{\partial \alpha}{\partial x} + \xi(\beta-\alpha),~~~~~~
\frac{\partial \beta}{\partial t} = v \frac{\partial \beta}{\partial x} - \xi(\beta-\alpha),
\end{equation}
where the total particle distribution yields $n(x,t) = \alpha(x,t)+\beta(x,t)$.
By performing simple mathematical operations on these equations, the following transport equation can be obtained
\begin{equation}\label{eq:alpha_plus_beta_rewritten}
\frac{\tau_\mathrm{s}}{2\epsilon}\frac{\partial^2 n}{\partial t^2} = \frac{v^2\,\tau_\mathrm{s}}{2\epsilon} \frac{\partial^2 n}{\partial x^2} - \frac{\partial n}{\partial t}.
\end{equation}
In fact, when we generalize this approach for three dimensions, assuming local homogeneity, this leads to the Telegraph equation in Eq.~(\ref{eq:telegraph}). Therefore, the statistics of particles that follow CRW agree with analytical theories of particle transport of cosmic rays \cite{Litvinenko2015, Tautz2016}.


\section{Comparison of Propagation Approaches in Astrophysical Environments}\label{sec:3}
In this section, we compare the performance of the propagation methods presented in Section~\ref{sec:2} for the problem described in the introduction: How do UHECRs propagate in the intergalactic magnetic field? Note that this question is only a piece of the puzzle of the complex question of the origin of these particles and serves in the following mainly as a realistic base case for which an analytical solution of the statistical propagation properties is known.

\subsection{Simulation Parameters}
The knowledge about the parameters of the IGMF still contains large uncertainties (see \cite{Alves2021} for a recent review). Taking current limits into account, we choose isotropic 3d Kolmogorov turbulence with a spectral index = $5/3$, a magnetic field strength $B_\mathrm{rms} = 1\,$nG, a correlation length $l_\mathrm{c} = 1\,$Mpc of the turbulence, and particles with $E = 10\,$EeV.

\subsection{Steady-State Diffusion Coefficients}
\begin{figure}[htbp]
    \centering
    \includegraphics[width=0.6\textwidth]{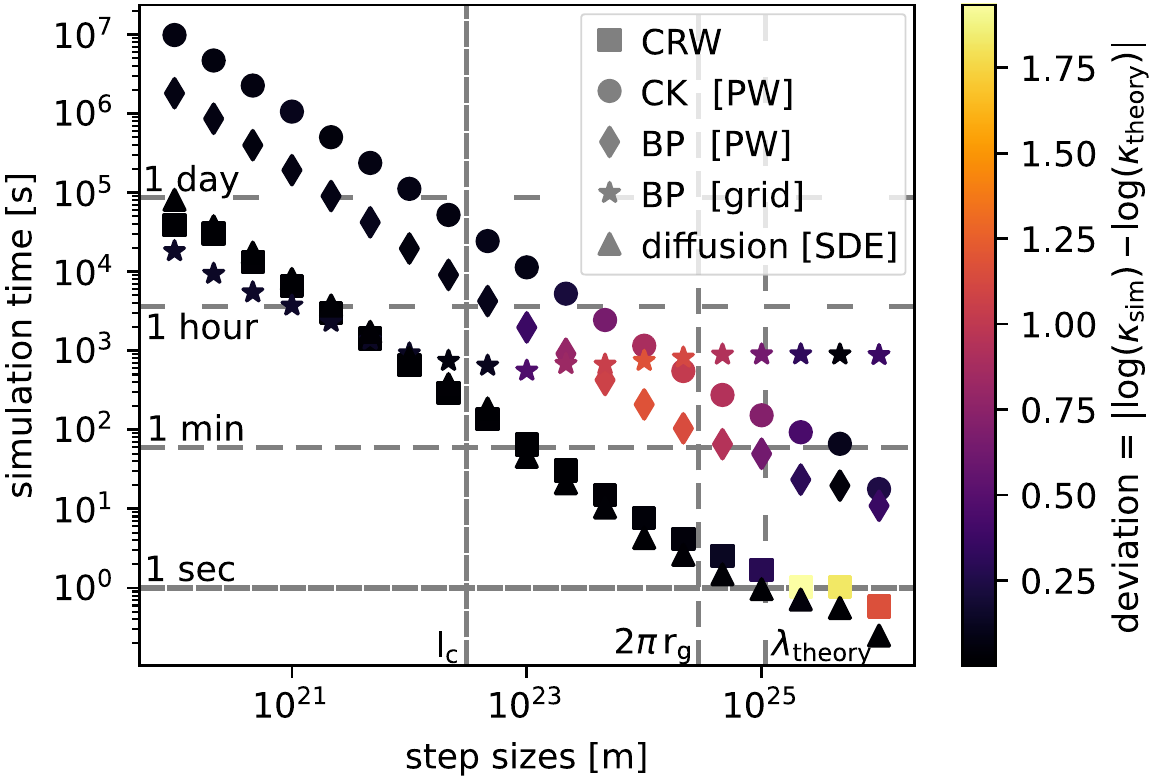}
    \caption{Comparison of simulation time (sum of the system and user CPU) for simulations with different propagation methods as functions of step sizes. Simulated with CRPropa 3.1.7 \cite{CRPropa3_1} and PropPy 1.0.0 \cite{reichherzer_patrick_2022_5959220}. Simulations can be reproduced via scripts available at \cite{reichherzer_patrick_2022_5959220} and shwon simulation results are available at \cite{reichherzer_patrick_2022_5959618}.
    \label{fig:fig2} 
    }
\end{figure}
In this simulation setup, the steady-state diffusion coefficients of the particles are determined using the various propagation methods and compared with the diffusion coefficient of $10^{33}\,$m$^2$/s expected from theory. Figure~\ref{fig:fig2} shows the results of the comparison by plotting the simulation times as a function of the step sizes used in the simulations. The deviations of the simulation results from the theory are color-coded. The diffusion approach gives very good agreement for every step size. The same is true for the CRW, with the exception of step sizes above the mean free path $\lambda_\mathrm{theory}$. Note that the CRW approach can additionally model the initial ballistic propagation phase well. Both methods are similarly fast. In contrast, the EOM methods are slow and only give good results when the step size is small enough to resolve the scales of gyration and fluctuations. The apparently improving agreement at large step sizes for EOM is only due to the superposition of two counteracting numerical errors.

EOM, CRW, and diffusive propagation methods are suitable for the propagation of charged particles in IGMF, as long as specific requirements are satisfied. Here, we have shown that this is especially true for step sizes and leads to severe limitations in EOM methods. In contrast to EOM methods, the diffusive and CRW methods are very fast. If small distances between source and observer are relevant, on whose scales particles do not become diffusive, CRW or EOM-based approaches should be preferred.

\end{document}